\documentclass[final,authoryear,3p,twocolumn]{elsarticle}

\usepackage{amssymb,hyperref}

\begin{document}


\def\aj{\rm{AJ}}                   
\def\aap{\rm{A\&A}}                
\def\mnras{\rm{MNRAS}}             
\def\apjs{\rm{ApJS}}               
\def\apjl{\rm{ApJ}}                
\def\apj{\rm{ApJ}}                 
\def\pasp{\rm{PASP}}               

\journal{Astronomy and Computing}


\begin{frontmatter}
\title{Galaxy detection and identification using deep learning and data augmentation.}


\author[myad1,myad2]{Roberto E. Gonz\'alez}
\address[myad1]{Centro I+D MetricArts, Santiago, Chile}
\address[myad2]{Centro de Astro-Ingenier\'{i}a, Pontificia Universidad Cat\'olica,
  Av. Vicu\~na Mackenna 4860, Santiago, Chile}
\ead{regonzar@astro.puc.cl}

\author[myad1]{Roberto P. Mu\~noz}
\author[myad2]{Cristian A. Hern\'andez}

\begin{abstract}
{We present a method for automatic detection and classification of galaxies which includes a novel data-augmentation
procedure to make trained models more robust against the data taken from  different instruments and contrast-stretching functions.
This method is shown as part of AstroCV, a growing open source computer vision repository for processing and analyzing big astronomical datasets, including high performance Python and C++ algorithms used in the areas of image processing and computer vision.
}


The underlying models were trained using convolutional neural networks and deep learning techniques, which provide better results than methods based on manual feature engineering and SVMs { in most of the cases where training datasets are large.}
The detection and classification methods were trained end-to-end using public
datasets such as the Sloan Digital Sky Survey (SDSS), the Galaxy Zoo, and private
datasets such as the Next Generation Virgo (NGVS) and Fornax (NGFS) surveys.

{ Training results are strongly bound to the conversion method from raw  FITS data for each band into a 3-channel color image. Therefore, we propose data augmentation for the training using 5 conversion methods. This greatly improves the overall galaxy detection and classification for images produced from different instruments, bands and data reduction procedures.}

The detection and classification methods were trained using the deep learning framework DARKNET and the real-time object detection system YOLO. These methods are implemented in C language and CUDA platform, and makes intensive use of graphical processing units (GPU). Using a single high-end Nvidia GPU card, it can process a SDSS image in 50 milliseconds and a DECam image in less than 3 seconds. 

We provide the open source code, documentation, pre-trained networks, python tutorials 
{, and how to train your own datasets, which can be found in the AstroCV repository. }
\href{https://github.com/astroCV/astroCV}{https://github.com/astroCV/astroCV}.

\end{abstract}


\end{frontmatter}


\section{Introduction}
Astronomical datasets are constantly increasing in size and complexity. The modern generation of integral field units (IFUs) are generating about $60$GB of data per night while imaging instruments are generating $300$GB per night. The Large Synoptic Survey Telescope \citep[LSST;][]{2008arXiv0805.2366I} is under construction in Chile and { it is expected to start full operations in 2022. With a wide 9.6 square degree field of view 3.2 Gigapixel camera, LSST will generate about $20$TB of data per night and will detect more than 20 million of galaxies.}

{ Machine learning techniques have been increasingly employed in data-rich areas of science. They have been used in genomics, high-energy physics and astronomy. Some examples in astronomy are the detection of weird galaxies using Random Forests on Sloan data \citep{2017MNRAS.465.4530B}, Gravity Spy \citep{0264-9381-34-6-064003} for LIGO detections and using convolutional neural network (CNN) in identifying strong lenses in imaging data \citep{2017MNRAS.471..167J}.}

\begin{table*}[!h]
\caption{\label{table1} Annotations in different subsamples.}
\begin{center}
{\small
\begin{tabular}{lrrrrr|c|c}  
\hline
\noalign{\smallskip}
Dataset & Elliptical & Spiral & Edge-on & DK & Merge & Total & Number images\\
\noalign{\smallskip}
\hline
\noalign{\smallskip}
Training S1 & 10366 & 4535 & 4598 & 223 & 381 & 20103 & 6458 \\
Validation S1 & 1261 & 714 & 723 & 27 & 45 & 2770 & 921 \\
Training S2 & 18030 & 7828 & 7910 & 350 & 648 & 34766 & 11010 \\
Validation S2 & 2119 & 856 & 873 & 36 & 82 & 3966 & 1161 \\
Custom & 705 & 401 & 462 & 474 & 135 & 2177 & 87\\
\end{tabular}
}
\end{center}
\end{table*}

Computer Vision is an interdisciplinary field that focuses in how machines can emulate the way in which human's brains and eyes work together to visually process the world around them. For many years, the detection of objects was computed using manual feature engineering and descriptors such as SIFT and HOG \citep{dalal_triggs_2005}. Thanks to the advent of large annotated datasets and gains in computing power, deep learning methods have become the favorite for doing detection and classification of objects.

The classification of optical galaxy morphologies is based on a few simple rules that make them suitable for machine learning and computer vision techniques. The Kaggle Galaxy Zoo \citep{2013MNRAS.435.2835W} was a competition based on a citizen science project where the aim was to predict the probability distribution of people's responses about the morphology of a galaxy using optical image data, and the winning solution used CNNs \citep{2015MNRAS.450.1441D}.


We present {a method for galaxy classification and identification with a novel data augmentation procedure which is part of }AstroCV, a computer vision library for processing and analyzing big astronomical datasets.
The goal of AstroCV is to provide a community
repository for fast Python and C++ implementations of common tools and routines used in the
areas of image processing and computer vision.
In particular, it is focused in the task of object detection, segmentation and classification applied to astronomical sources.

In this paper we will focus in the automatic detection and classification of galaxies.
The detection and classification methods were trained end-to-end using public
datasets from the Sloan Digital Sky Survey (SDSS), \citep{2015ApJS..219...12A},
and Galaxy Zoo \citep{2008MNRAS.389.1179L,2011MNRAS.410..166L} explained in section $2.1$
We use YOLO method, \citep{2015arXiv150602640R}, for object detection which is explained in section $2.2$.
Training process is described in section $2.3$, and results are shown in section $3$.

The open source code, training datasets, documentation and python notebooks of AstroCV are freely available in a Github repository\footnote{\href{https://github.com/astroCV}{https://github.com/astroCV}.}.

\section{Data and Training}

\subsection{Dataset}

Galaxy Zoo \footnote{\href{https://www.galaxyzoo.org/}{https://www.galaxyzoo.org/}.} \citep{2008MNRAS.389.1179L, 2011MNRAS.410..166L} is the most successful citizen project in Astronomy. It consists of a web platform for doing visual inspection of astronomical images and morphological classification of galaxies. Hundreds of thousands of volunteers classified images of nearly $900,000$ galaxies drawn from the SDSS survey. The Galaxy Zoo classification consists of six categories: elliptical, clockwise spiral, anticlockwise spiral, edge-on, star/do not know, or merger.

 We extracted the galaxy classification for a sub-sample of 
$38,732$ galaxies and downloaded their respective gri-band images from the SDSS fields.
Sub-sample S1 is produced from $20000$ field images and sub-sample S2 is produced from $32000$ field images.

For each field image, we select the galaxies with a size larger than $22$ pixels box side. This galaxy size is computed as $2.1$ times the $r$-band petrosian radius. 
{
After this size filter we stay with two samples of $7397$ images and $12171$ images. Then,
we split each of these two samples into training and validation sub-sets, resulting in S1 and S2 datasets.
In addition, we include a small custom dataset (Custom hereafter)with manually annotated galaxies from Hubble Deep Field image gallery
\footnote{\href{http://www.spacetelescope.org/science/deep_fields}{http://www.spacetelescope.org/science/deep$\_$fields}.}, CFHT\footnote{\href{http://www.cfht.hawaii.edu}{http://www.cfht.hawaii.edu}}, and others images randomly taken from public databases. 
}
See \autoref{table1} with details of the different samples.

\subsection{YOLO}

\begin{figure*}[!ht]
\begin{center}
\includegraphics[width=.99\linewidth]{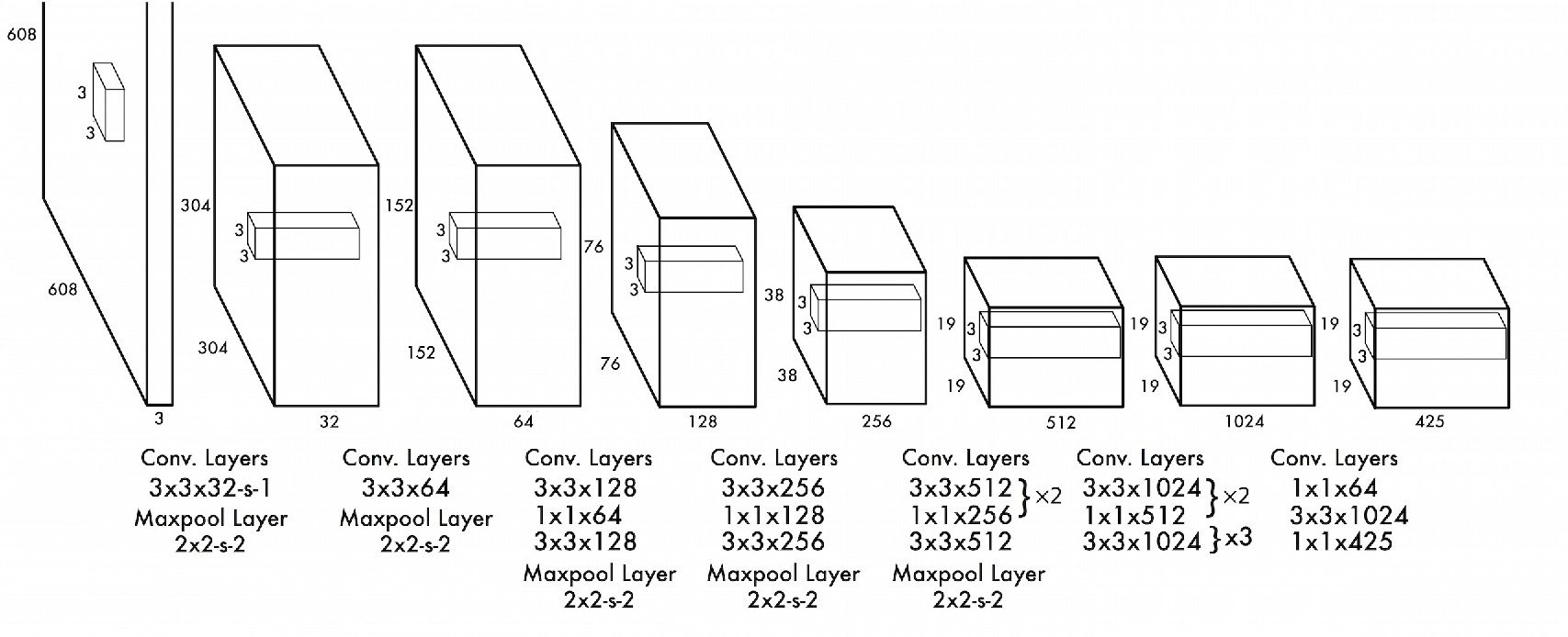}
\caption{
\label{fig3ref} YOLO network schema.}
\end{center}
\end{figure*}

You only look once (YOLO) method \citep{2015arXiv150602640R,2016arXiv161208242R}, is a Single Shot Detector (SSD),
it means it computes in a single network the region proposal and classifier. This method run the image on a Convolutional Neural Network (CNN) model and get the detection on a single pass. The network is composed with $23$ convolution layers and $5$ maxpool layers shown in \autoref{fig3ref}, and it is programmed on Darknet, an open source neural network framework in C and CUDA. It is very fast and take full advantage of graphical processing units (GPU).
This method formerly developed for person/object detection, is configured for the training and detection of galaxies.

\subsection{Training and data Augmentation}

YOLO method is designed to work on 3 channel color images, usually RGB color scale. In astronomy images are taken for each filter in FITS format with raw CCD data for each pixel.
Data conversion from FITS to RGB images(or contrast stretching) is not unique and depends on the telescope's camera, band filters, reduction schema, and most important, it depends on the conversion method used to scale photon counts to color scale. 

There are several conversion methods, however to emphasize galaxies with strong luminosity gradients, linear scaling is not suitable, i.e. a spiral galaxy radial luminosity profile can be modeled as a power law(de Vaucouleurs profile) for the bulge plus an exponential for the disk. In those cases, the scaling methods commonly used are $sinh$, $asinh$, $sqrt$ functions.
SDSS use \citet{2004PASP..116..133L} as standard conversion method from FITS in $igr$ bands to RGB an image (Lupton method hereafter).

In general, to train neural networks using images, it is fundamental the data augmentation, it means increase the training dataset with different transformations of the same data (scaling, rotations, crop, warp, etc); In YOLO this data augmentation is already implemented in the training procedure, however we need to produce a color-scale/filter conversion augmentation as well (filter hereafter), to build a training robust against RGB images coming from different filters, bands and instruments.

{ 
In the top performance deep learning methods for object detection, we have also Faster R-CNN, Single shot detectors(SSD), Deconvolutional DSSD, Focal Loss, Feature Pyramid Networks (FPN). In \citet{2017arXiv170802002L} there is a complete review and comparison on current methods. Most of these methods present similar mean average precision when compared to YOLO, however we stay with the latter since it is the fastest and implemented in C with CUDA acceleration.
}

\begin{table}[!htb]
\caption{\label{table2} Training sets.}
\smallskip
\begin{center}
{\small
\begin{tabular}{l|c|c|c}  
\hline
\noalign{\smallskip}
Name & Dataset & Filters & Images\\
\noalign{\smallskip}
\hline
\noalign{\smallskip}
T1 & S1 & L & 6458\\
T2 & S1 & LH& 6458\\
T3 & S2 & L& 11010\\
T4 & S2 & L+LH+S+SH+Q& 55050\\
T5 & S2 & LH+SH & 22020\\
T6 & S2 & L+LH+S+SH+Q & 32290 \\
\hline
\end{tabular}
}
\smallskip
\raggedright \footnotesize{(L=Lupton, LH=Lupton high contrast, S=sinh, SH=sinh high contrast, Q=sqrt; C=custom Hubble sample.)}
\end{center}
\end{table}

{
In \autoref{table2} we show 5 different training sets with RGB images produced to explore dataset size and filter augmentation, we use $Lupton$, $Lupton$ high contrast\footnote{Contrast and Brightness enhanced by $2$ using ImageEnhance python library.}, $sinh$, $sinh$ high contrast, { and $sqrt$ conversion functions.} 
}

The purpose of these 5 training sets(T1,T2,T3,T4,T5 hereafter) is to test color scale conversion effects, dataset size,  and produce a training robust against any astronomical FITS image converted to RGB.
{T1 and T2 are intended to show training differences produced between two similar contrast strectching functions; T3 will show the effect of increasing dataset size when compared to T1; The effect of including additional contrast stretching filters for data augmentation will be obtained from comparison between T3, T5 and T6. Finally T6, will be used to check the effect of dataset size in the case of augmented trainings when compared with T4.}


\begin{figure*}[!thb]
\begin{center}
\includegraphics[width=.96\linewidth]{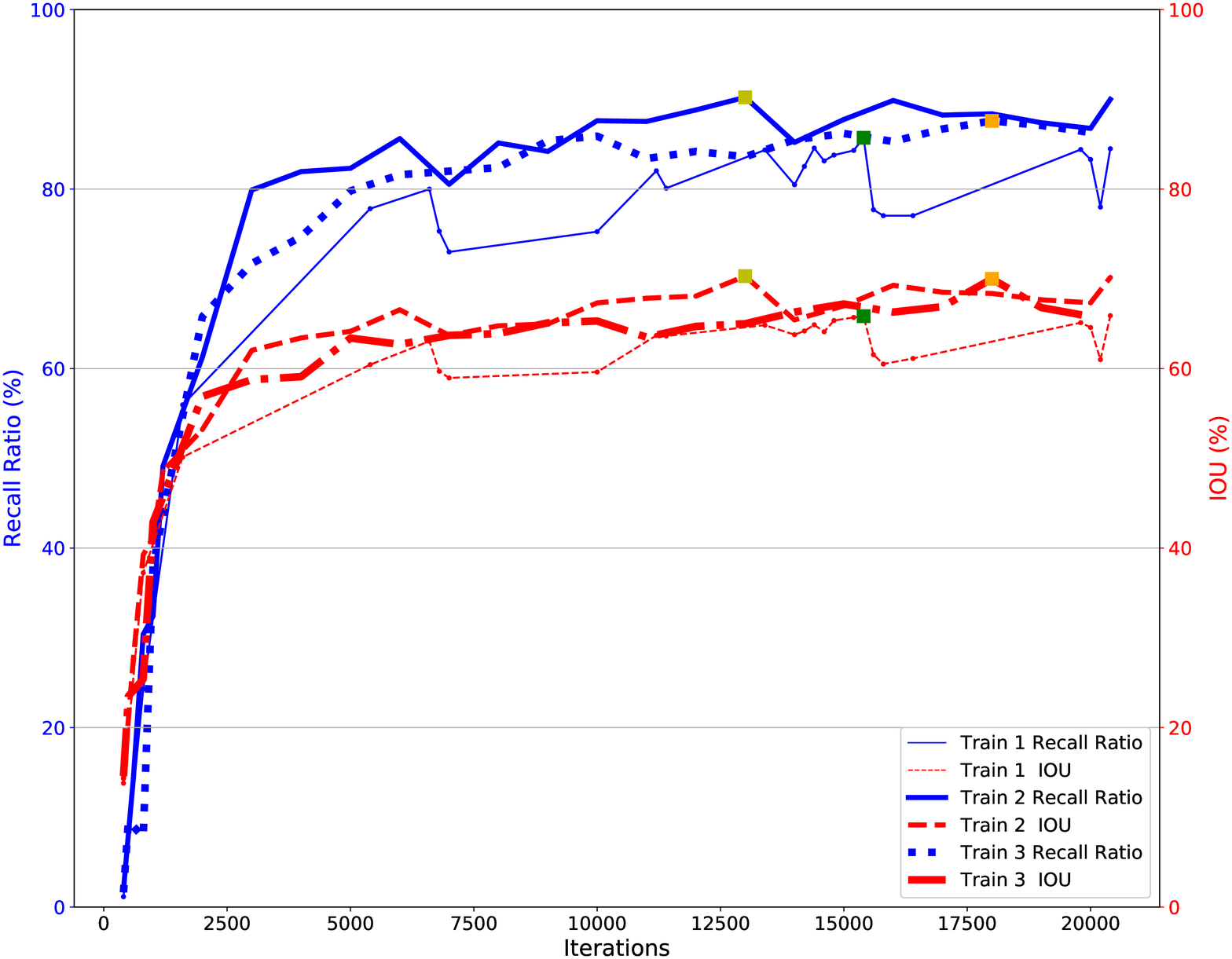}
\caption{\label{fig1ref} Convergence of T1, T2 and T3 training sets. Green, yellow and orange squares indicate optimal recall ratio and IOU for the three trainings. Comparison between traditional(T1) and high contrast(T2) Lupton filters show a detection improvement of $5\%$ for T2. Increasing the dataset size(T3) for lupton filter(T1) show a detection improvement of $2\%$.}
\end{center}
\end{figure*}

\begin{figure*}[!thb]
\begin{center}
\includegraphics[width=.96\linewidth]{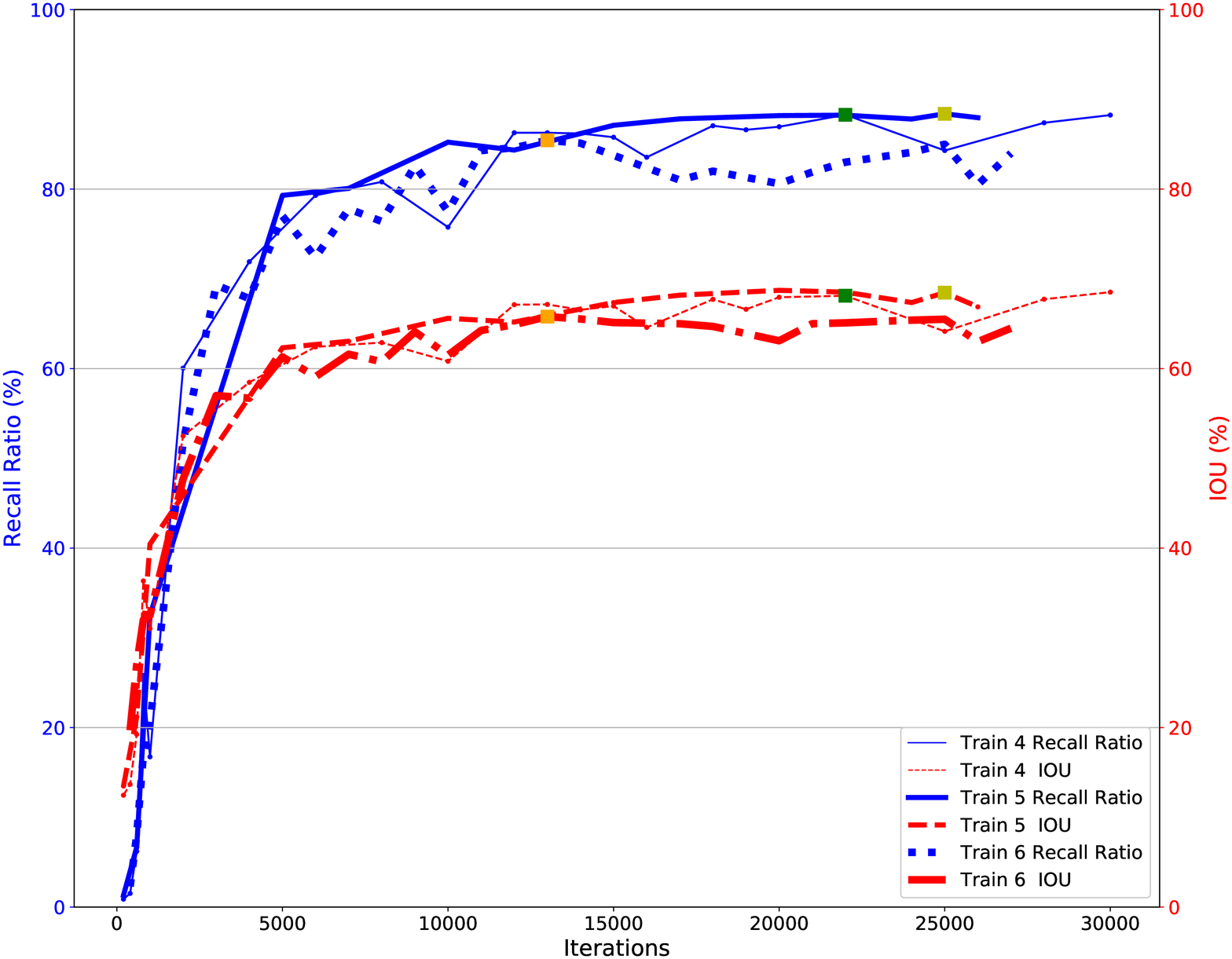}
\caption{\label{fig5ref} Convergence of the T4, T5 and T6 training sets where data is augmented by several conversion filters. Recall and IOU remain similar to \autoref{fig1ref}, however this time detection and classification is more robust against different contrast stretching functions, instruments or reduction processes.}
\end{center}
\end{figure*}

\section{Results}

In the training process, we look for the optimal number of iterations where the network accuracy converges.
We look for maximum recall ratio (fraction of detections recovered from the ground thruth), and IOU (Intersect over union, overlap fraction between the detected and ground thruth bounding box) {running the training over the corresponding validation sets.}

We can explore the effect of brightness and contrast using Lupton filter by comparing T1 and T2, results are shown in \autoref{fig1ref},
where we get for T1, a maximum recall ratio and IOU of $85.7\%$ and $65.8\%$ respectively. In the case of T2, maximum recall ratio and IOU rise to $90.23\%$ and $70.3\%$ respectively.
T2 returns better results since higher contrast and brightness images built from FITS images gives more information to the training, in particular for disk galaxies where disk can be seen more clearly with a higher contrast and brightness. However, excessive brightness may add too much noise to the background and excessive contrast tend to erase central bulge profile.
{In \autoref{fig1ref} we also show T3 where we get $87.6\%$ and $69.8\%$ maximum recall and IOU, when compared with T1, we  have that increasing the dataset size for a fixed conversion filter improves results accuracy, however this improvement is smaller than the case of using a different high contrast conversion filter. }
 
We also explore data augmentation using additional color conversion filters. As shown in \autoref{table2} 
{T4 and T6 are augmented using $5$ filters and T5 using $2$ filters.}
\autoref{fig5ref} show results for T4, T5 and T6, where we get for T4, maximum recall ratio and IOU of $88.28\%$ and $68.12\%$ respectively, in the case of T5, maximum recall ratio and IOU are $88.24\%$ and $68.52\%$ respectively,
{and in the case of T6, maximum recall ratio and IOU are $85.44\%$ and $65.79\%$ respectively.} 
Detection and classification accuracy remains close to $90\%$ regardless we are using data augmentation with 
{ $2$ or $5$ filters,  or if we are using the S1 or S2 sample.}
This allows the method to be robust against a wide range of color scale and filter conversions from FITS to RGB images,
{ without losing detector accuracy.}

{
Regarding data sample size, we explore dataset size effect on training using S1 and S2. As shown in \autoref{table1} dataset S2 is $75\%$ larger than dataset S1, in addition, 
when we compare recall ratio and IOU converge between S1 and S2 based trainings(for instance, T1 and T3, or T4 and T6),
we have that S2 based trainings show a slightly better accuracy and smoother convergence compared to S1 based training sets. From this result, we can expect that using a larger training set should not improve substantially method accuracy/convergence and we are close to YOLO network maximum accuracy for this particular model. 
}

\begin{table*}[!ht]
\caption{\label{table3} Confusion matrix for galaxy classification using T2. Predicted values(columns) vs actual values(rows).}
\smallskip
\begin{center}
{\small
\begin{tabular}{lrrrrr|c}  
\hline
\noalign{\smallskip}
n=2756 & Elliptical & Spiral & Edge-on & DK & Merge & Recall\\
\noalign{\smallskip}
\hline
\noalign{\smallskip}
Elliptical & 1172 & 33 & 57 & 1 & 2 & 0.92 \\
Spiral & 176 & 469 & 69 & 0 & 3 & 0.65 \\
Edge-on & 96 & 60 & 554 & 0 & 3 & 0.78 \\
DK & 6 & 9 & 3 & 3 & 0 & 0.14 \\
Merge & 26 & 2 & 3 & 0 & 9 & 0.23 \\
\hline
Precision & 0.79 & 0.82 & 0.81 & 0.75 & 0.53 & Accuracy=0.80 \\
\end{tabular}
}
\end{center}
\end{table*}

\begin{table*}[!ht]
\caption{\label{table4} Confusion matrix for galaxy classification using T4. Predicted values(columns) vs actual values(rows).}
\smallskip
\begin{center}
{\small
\begin{tabular}{lrrrrr|c}  
\hline
\noalign{\smallskip}
n=19246 & Elliptical & Spiral & Edge-on & DK & Merge & Recall\\
\noalign{\smallskip}
\hline
\noalign{\smallskip}
Elliptical & 9525 & 476 & 257 & 0 & 3 & 0.93 \\
Spiral & 940 & 2991 & 305 & 0 & 6 & 0.71 \\
Edge-on & 688 & 499 & 3063 & 0 & 4 & 0.72 \\
DK & 39 & 36 & 24 & 6 & 4 & 0.06 \\
Merge & 170 & 129 & 44 & 0 & 37 & 0.10 \\
\hline
Precision & 0.84 & 0.72 & 0.83 & 1.0 & 0.69 & Accuracy=0.81 \\
\end{tabular}
}
\end{center}
\end{table*}

{
We explore in more detail the two most relevant training sets T2 and T4. T2 have the highest recall ratio { using a single conversion filter} and T4 includes the largest data augmentation.

To study the accuracy of the method we compute the confusion matrix for  T2 and T4 in \autoref{table3} and \autoref{table4} respectively. 
Te confusion matrix is constructed using the corresponding validation sets for T2 and T4 with $2770$ and $3966\times5$ galaxies respectively.
We use a detection threshold of $15\%$. Columns indicate predicted values and row the ground truth\footnote{Notice that ground truth in the context of galaxy zoo define the galaxy class based on the number of votes.}. In the top left corner we show the number of detected galaxies, notice that numbers does not match exactly with the validation set from \autoref{table1}, this is because many reasons, such as, merge objects detected as two objects; other reason is the detection threshold value; and a third reason is that detection process includes non-maximum suppression(NMS) issues, this correspond to the post-processing algorithm responsible for merging all detections that belong to the same object or overlapping objects. However, the sum of all theses effects is small, in both training set we miss less than $3\%$ detections. 

Confusion matrix for both trainings show very little confusion between the three major classes elliptical, spiral and edge-on. In the case of DK and Merge classes the number of annotation is very small compared to the other classes resulting in poor performance for those classes. In addition, merge class in galaxy zoo is very ambiguous since the thin line separating a blended galaxy pair and two independent close galaxies.

The total classification accuracy is $80\%$ for T2 and $81\%$ for T4, which is high taking into account the nature of Galaxy Zoo classification based on votes, where some objects have divided votes but are assigned to a single classification (i.e. Dwarf spheroidal or S0 galaxies). This show that conversion filter data augmentation does not degrade method accuracy, and make the training more robust against different image sources comming from different instruments.}


\begin{figure*}[!htb]
\begin{center}
\includegraphics[width=.99\linewidth]{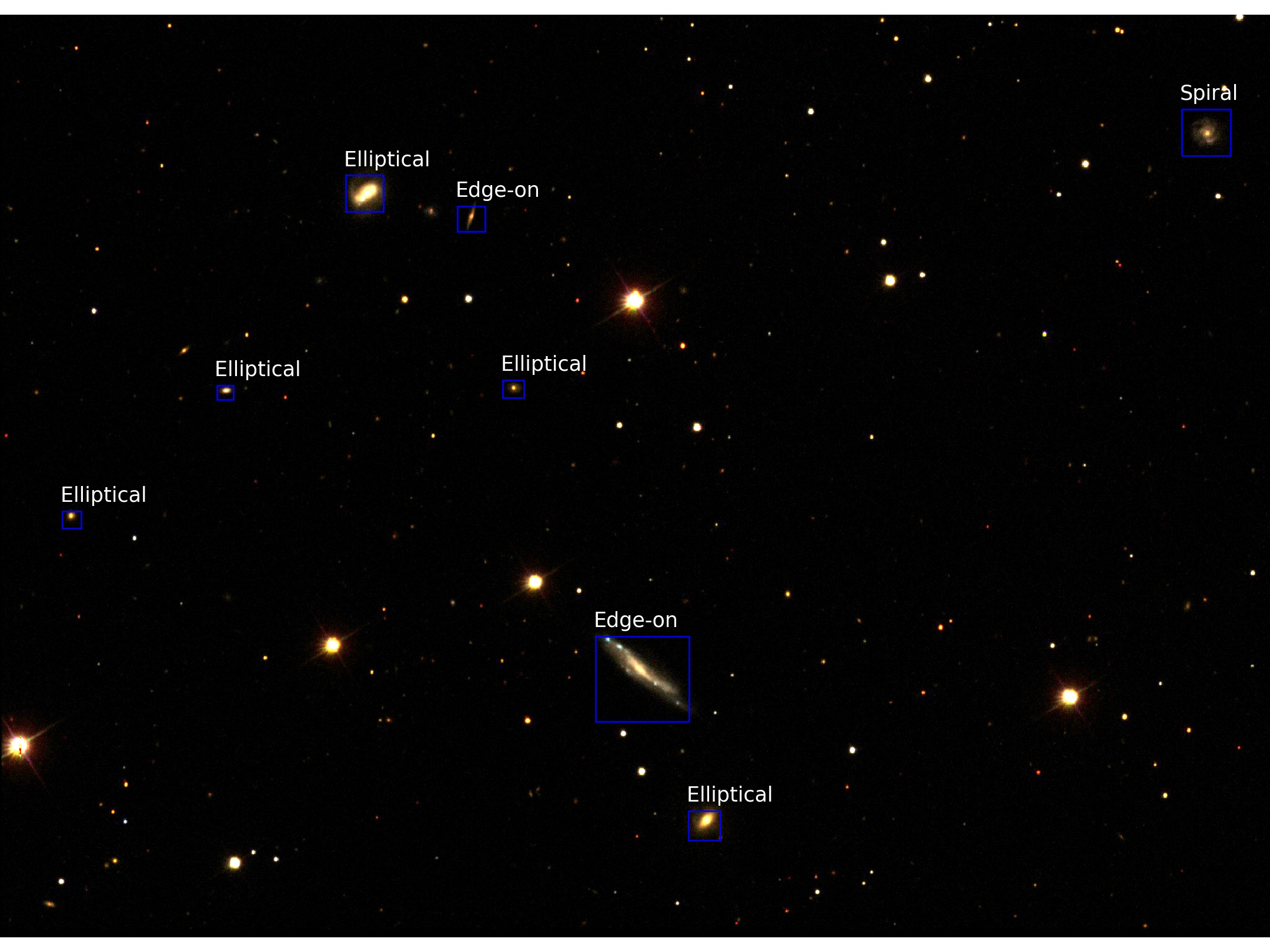}
\caption{
\label{fig2ref} Detection and Classification of galaxies in a $2048x1489$ image from SDSS using T2 training; i, r, and g filters converted to rgb color space using Lupton filter.}
\end{center}
\end{figure*}

\begin{figure*}[!htb]
\begin{center}
\includegraphics[width=.70\linewidth]{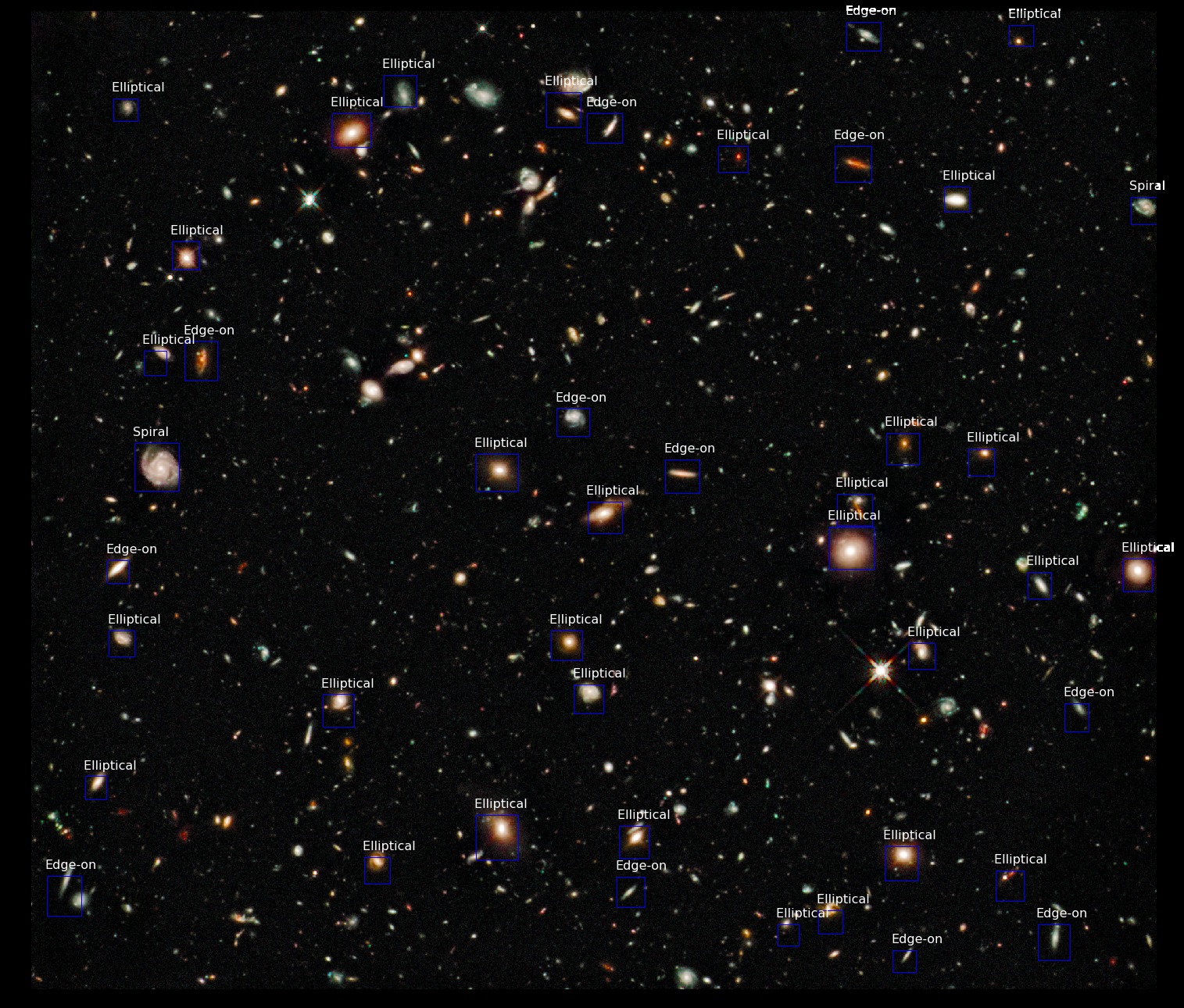}

\vspace{0.2cm}

\includegraphics[width=.70\linewidth]{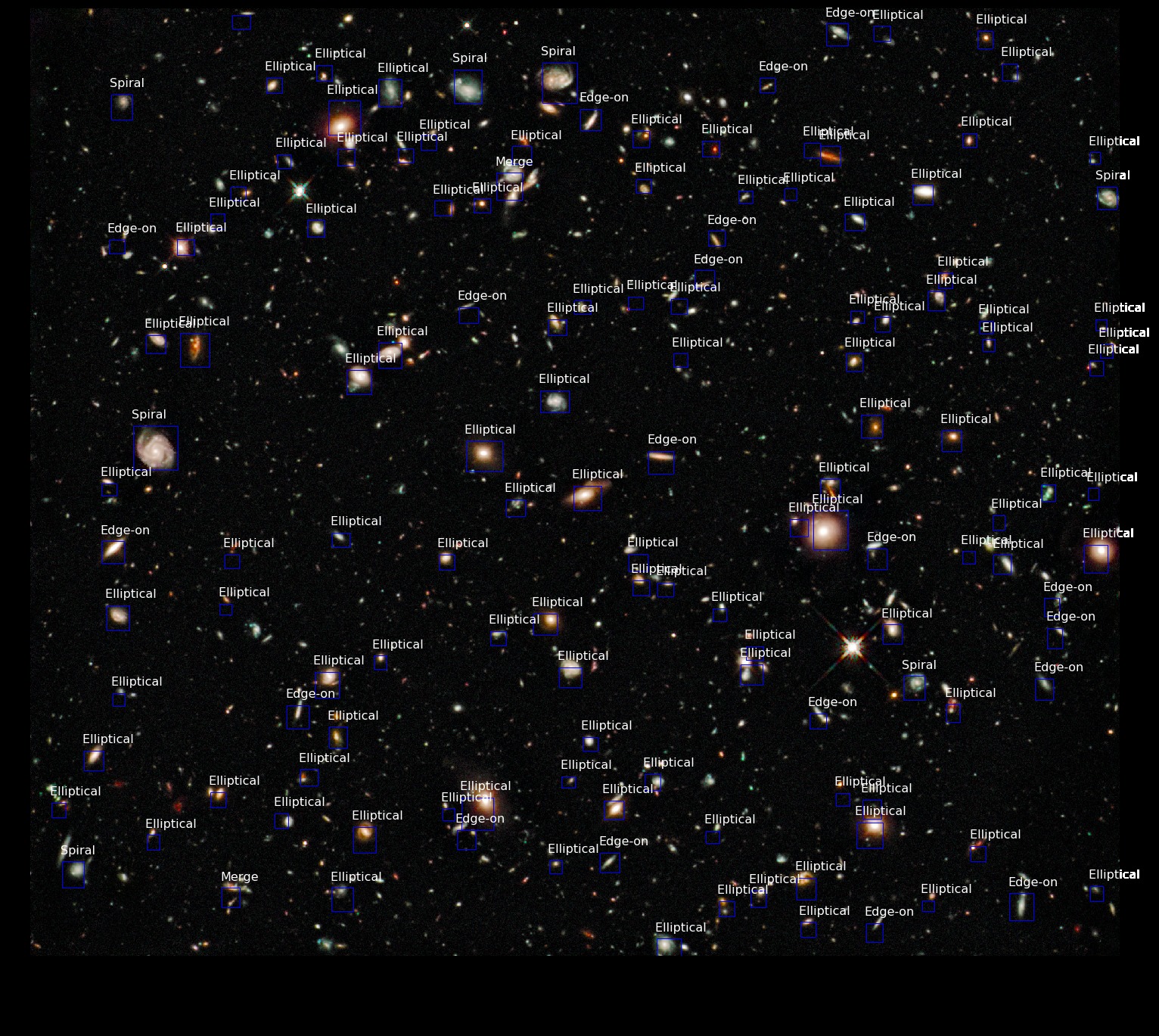}
\caption{
\label{fig4ref}
{ Galaxy detection results on a Hubble deep field image. Top panel: Using T2(44 detections). Bottom panel: Using augmented training T4(140 detections). }
}
\end{center}
\end{figure*}

{
In \autoref{fig2ref}, we show results of the method using T2 on an image from the SDSS converted using Lupton filter. In the github repository there are examples to reproduce this figure.

In \autoref{fig4ref} we show result over a Hubble Deep field image(taken from Custom dataset) using T2 and T4. This image use infrared bands and use a different color scale conversion, not comparable with SDSS filters and conversions. However, we can see how filter data augmentation from T4 produces a $3x$ improvement in the detection of galaxies compared with T2. Then T4 training is more robust to images produced from different filters and instruments.

{
To quantify the improvement of data augmentation for the different trainings, we validate them using the custom dataset which include images randomly taken from different telescopes and filtes, and manually annotated. Results for the recall ratio and IOU are shown in \autoref{table5}. Performance comparison between (T1,T2,T3) and (T4,T6), show that data augmentation using 5 contrast stretching functions, improve detection and identification of galaxies in random images by a factor of $2-3$. One to one comparison between trainings indicates: i) T1-T2, LH gives more information to the training compared to standard Lupton conversion; ii) T1-T3 or T4-T6, increasing the dataset may improve detector accuracy, but just a little when compared to data augmentation; iii) T3-T5-T4, there is a consistent increasingly improvement in detection accuracy when including more data augmentation of the same data. 
}

}

\begin{table}[!htb]
\caption{\label{table5} Detection accuracy for different training sets validated with the custom dataset.}
\smallskip
\begin{center}
{\small
\begin{tabular}{l|c|c}  
\hline
\noalign{\smallskip}
Training Set & Recall Ratio  & IOU Ratio\\
\noalign{\smallskip}
\hline
\noalign{\smallskip}
T1 & 0.120 & 0.236 \\
T2 & 0.183 & 0.296\\
T3 & 0.192 & 0.292\\
T4 & 0.404 & 0.423\\
T5 & 0.271 & 0.336\\
T6 & 0.364 & 0.403\\
\hline
\end{tabular}
}
\end{center}
\end{table}


\section{Discussion}


The current classification model was created using a sample of SDSS galaxies from the Galaxy Zoo Project.
{ We propose a data augmentation based on different additional filters to convert FITS data from different bands into RGB color scale, and this show an important improvement in comparison with traditional training using a single color scale conversion, specially with images produced with bands and conversion filters totally different from the used in the former training(see \autoref{fig4ref} and \autoref{table5}). }
{ We show a factor of $\sim3$ improvement in the detection and classification of random galaxy images(Custom dataset) taken from different telescopes, bands, and contrast stretching functions.}
This makes the training set more robust and generic to be applied to images produced from any telescope and band
{, and show that increasing the dataset size for training is not necessarily the most important target to produce a good galaxy detector/classificator, while a well based data augmentation may lead to more significant improvements. }

{
This method is part of the AstroCV library intended to provide computer vision and image processing tools for processing big astronomical datasets. These tools are focused in the task of object detection, segmentation and classification of astronomical objects using deep learning and neural network frameworks.
This method is implemented on DARKNET which take advantage of GPUs capabilities, improving performance for real-time applications such as data reduction pipelines. As an example, running galaxy detection using a Titan-X Nvidia GPU card on a typical full HD image takes nearly $50$ms, and for a DECAM $500$Mpx images it takes less than three seconds.
}

There are several other interesting datasets to be trained such as Galaxy Zoo 2, \citep{2013MNRAS.435.2835W} and the Next Generation Fornax Survey \citep[NGFS;][]{2015ApJ...813L..15M}. The Galaxy Zoo 2 extends the classification for nearly $300,000$ of the largest Galaxy Zoo 1 galaxies, including detailed features such as discs, bars, and arms. The NGFS survey extends the classification to low surface brightness galaxies down to $\mu_i=28\; \rm{mag}\, \rm{arcsec}^2$.

The next generation of astronomical observatories, such as the Large Synoptic Survey Telescope (LSST) and the Extreme Large Telescope (ELT), will observe hundreds of thousands of galaxies and will generate Terabytes of data every night. We are planning to improve the scalability of AstroCV to do real-time processing of tomorrow's big astronomical datasets. Another interesting applications we are exploring is the identification and classification of transient objects, { and detection of high redshift candidates for spectroscopy}.

{ In this paper we limit the training procedure for three color bands only, to produce the RGB contrast stretching, since this is the classic approach to manually find features and classify galaxies (i.e. Galaxy Zoo). However, training may be strongly improved if we use more color bands(i,e. ugriz), or an optimal contrast stretching, and this will be part of further work.}

{
AstroCV is open source, the github repository at \href{https://github.com/astroCV/astroCV}{https://github.com/astroCV/astroCV} include code, documentation, pre-trained networks, python tutorials for using the library  and train your own datasets.
 }

\section*{Acknowledgements}

The authors acknowledges support from the NVIDIA developer program. The Centro I+D MetricArts acknowledges support from the NVIDIA Inception program, NVIDIA GPU Grant Program and the Microsoft BizSpark Program. We are grateful to Patricio Cofre, Wilson Pais and Adrian Fernandez for helpful discussions.
RG was supported by Proyecto Financiamiento Basal PFB-06 'Centro de
Astronomia y Tecnologias Afines'.

Funding for SDSS-III has been provided by the Alfred P. Sloan Foundation, the Participating Institutions, the National Science Foundation, and the U.S. Department of Energy Office of Science. The SDSS-III web site is \href{http://www.sdss3.org/}{http://www.sdss3.org/}.

\InputIfFileExists{astrocv.bbl}



\end{document}